\documentclass[prl,twocolumn]{revtex4-2}
\usepackage[utf8]{inputenc}
\usepackage[T1]{fontenc}
\usepackage{amsmath}
\usepackage{amsfonts}
\usepackage{amssymb}
\usepackage{graphicx}
\usepackage{xcolor}
\usepackage{hyperref}
\usepackage{comment}

\begin{document}

\title{Quantum and classical spin network algorithms \\ for $q$-deformed Kogut-Susskind gauge theories}

	\author{Torsten V. Zache}
    \email{torsten.zache@uibk.ac.at}
    \author{Daniel Gonz\'{a}lez-Cuadra}
    \author{Peter Zoller}

\affiliation{Institute for Theoretical Physics, University of Innsbruck, 6020 Innsbruck, Austria}
\affiliation{Institute for Quantum Optics and Quantum Information of the Austrian Academy of Sciences,
6020 Innsbruck, Austria}
	
	\begin{abstract}
        Treating the infinite-dimensional Hilbert space of non-abelian gauge theories is an outstanding challenge for classical and quantum simulations.
        Here, we introduce \emph{$q$-deformed Kogut-Susskind lattice gauge theories}, obtained by deforming the defining symmetry algebra to a quantum group.
        In contrast to other formulations, our proposal simultaneously provides a controlled regularization of the infinite-dimensional local Hilbert space while preserving essential symmetry-related properties. 
        This enables the development of both quantum as well as quantum-inspired classical \textbf{S}pin \textbf{N}etwork \textbf{A}lgorithms for \textbf{Q}-deformed gauge theories (\emph{SNAQs}). 
        To be explicit, we focus on SU(2)$_k$ gauge theories, that are controlled by the deformation parameter $k$ and converge to the standard SU(2) Kogut-Susskind model as $k \rightarrow \infty$.
        In particular, we demonstrate that this formulation is well suited for efficient tensor network representations by variational ground-state simulations in 2D, providing first evidence that the continuum limit can be reached with $k = \mathcal{O}(10)$. Finally, we develop a scalable quantum algorithm for Trotterized real-time evolution by analytically diagonalizing the SU(2)$_k$ plaquette interactions.
        Our work gives a new perspective for the application of tensor network methods to high-energy physics and paves the way for quantum simulations of non-abelian gauge theories far from equilibrium where no other methods are currently available.
	\end{abstract}
\maketitle

\paragraph{Introduction.--}
    Lattice gauge theories (LGTs) constitute the foundation of our fundamental understanding of nature, as formulated in the Standard Model of particle physics~\cite{montvay1994quantum}, as well as the spin foam approach to quantum gravity~\cite{rovelli2015covariant}. LGTs also find applications in the study of topologically ordered phases in condensed matter physics~\cite{levin2005string} and quantum information processing~\cite{kitaev2003fault}.
    The lattice formulation~\cite{wilson1974confinement,wegner1971duality,kogut1979introduction}, discretizing space and time while preserving the relevant symmetries of the theory, allowed to put gauge theories on a computer, eventually leading to remarkable predictions in QCD~\cite{davies2004high}. These well-established methods are, however, hindered by numerical sign problems~\cite{troyer2005computational} that arise, e.g., for real-time dynamics or in the presence of fermionic matter.
    
    In recent years, quantum-inspired classical methods, such as tensor networks that target physically relevant low-entangled states~\cite{cirac2021matrix}, have emerged as promising alternatives to simulate LGTs without sign problems~\cite{banuls2020review,meurice2022tensor,montangero2022loop}. On the other hand, quantum computers and simulators can more efficiently tackle highly-entangled regimes~\cite{banuls2020simulating,aidelsburger2022cold,zohar2022quantum,wiese2022quantum,klco2018quantum}, and we refer to \cite{martinez2016real,schweizer2019floquet, kokail2019self, mil2020scalable,yang2020observation,klco20202,zhou2022thermalization, nguyen2022digital,mildenberger2022probing,frolian2022realizing,atas20212,atas2022real} for experimental realizations of LGTs.
    While the simulation of non-abelian LGTs is arguably one of the most promising targets for a potential quantum advantage~\cite{daley2022practical},
    treating the infinite-dimensional Hilbert space of non-abelian theories remains an outstanding theoretical challenge~\cite{byrnes2006simulating,zohar2015formulation,zohar2017digital,lamm2019general,wiese2022quantum,liu2013exact,ciavarella2021trailhead,mathur2005harmonic,raychowdhury2020loop,liu2022qubit,kreshchuk2022quantum,cunningham2020tensor,jakobs2023canonical} and previous approaches have suffered from fundamental drawbacks.
    In particular: (i) finite subgroup truncations~\cite{zohar2017digital,lamm2019general} ultimately lead to uncontrolled errors because any non-abelian Lie group has a largest finite subgroup; (ii) quantum link models~\cite{wiese2022quantum} give up unitarity of the plaquette operator,
    rendering known efficient decompositions inapplicable;
    (iii) hard cutoffs in the ``representation'' basis~\cite{tagliacozzo2014tensor,ciavarella2021trailhead,tong2022provably,davoudi2022general} typically require more sophisticated quantum algorithms as subroutines leading to hardware requirements beyond the realm of current ``Noise Intermediate-Scale Quantum'' (NISQ) devices. For a recent comparison of different Hamiltonian formulations of LGTs, we refer to~\cite{davoudi2021search}.

    In this letter, we propose to overcome these problems with a new LGT formulation, which is tailored for quantum algorithms but also serves as a natural starting point for quantum-inspired classical methods.
    In addition to the spatial lattice regularization underlying the Kogut-Susskind (KS) formulation~\cite{kogut1975hamiltonian}, we regularize the infinite-dimensional Hilbert space resulting from non-abelian Lie groups by replacing the corresponding Lie algebra with a quantum group~\cite{biedenharn1995quantum}. 
    In a basis of gauge-invariant spin network (SN) states, we thus define a truncated model, which we call \emph{$q$-deformed Kogut-Susskind (qKS) LGT}, and argue that it preserves essential symmetry-related properties of the model, while the KS theory is recovered by tuning a single control parameter $k \in \mathbb{N}$~\footnote{The process of replacing a Lie algebra $\mathfrak{g}$ with a quantum group $\mathcal{U}_q \left(\mathfrak{g}\right)$ is often referred to as a ``$q$-deformation''~\cite{biedenharn1995quantum}. In our case, where $\mathfrak{g} = \mathfrak{su}(2)$, the deformation parameter is given by $q=e^{2\pi i /(k+2)}$.}. 
    
    Here, we study the case of SU(2)$_k$ LGT in two spatial dimensions in detail and first show the convergence of the $k \rightarrow \infty$ limit with exact results for a single plaquette.
    We then illustrate the advantages of this formulation by developing both classical and quantum \textbf{S}pin \textbf{N}etwork \textbf{A}lgorithms for \textbf{Q}-deformed gauge theories (\emph{SNAQs}).
    In the classical case, we perform tensor network simulations based on a simple iPEPS~\cite{cirac2021matrix} ansatz, indicating quantitative agreement with continuum results for $k = \mathcal{O}(10)$. 
    Concerning quantum simulations, we design a scalable digital quantum algorithm for real-time evolution using an analytical Trotter decomposition, which is enabled
    by an exact diagonalization of the plaquette operator using local basis transformations on a SN register.

\begin{figure}
    \centering
    \includegraphics[width=\columnwidth]{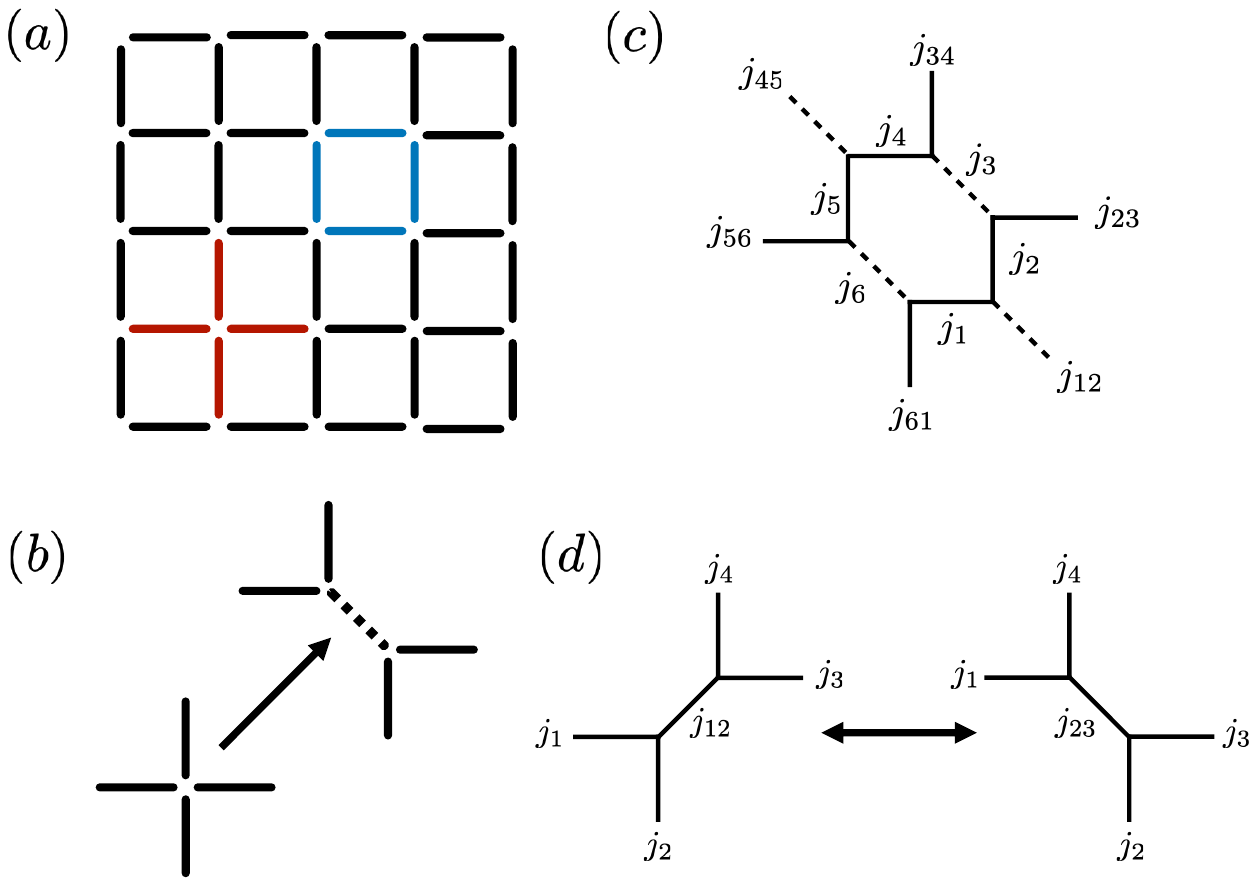}
    \caption{(a) In a 2+1D Kogut-Susskind LGT [see Eq.~\eqref{eq:KS}] gauge fields live on links of a spatial 2D square lattice, which contains elementary plaquettes (blue) and four-vertices (red). (b) For the gauge-invariant SN basis (see main text) every four-vertex is split into two three-vertices, resulting in an additional link (dashed). (c) The elementary plaquette operator on the point-split lattice acts on elementary hexagons according to Eq.~\eqref{eq:plaquette}. (d) A key feature preserved by our proposed $q$-deformed regularization are local unitary transformations (``$F$-moves'') that effect a basis transformation between inequivalent ways of point-splitting [see Eq.~\eqref{eq:F_operator}].}
    \label{fig1}
\end{figure}

\paragraph{Model and truncation.--}
To be specific, we consider a SU(2) LGT in two spatial dimensions, but our approach applies to all SU(N) LGTs in arbitrary dimensions. In preparation for the $q$-deformed theory, we start with the KS Hamiltonian~\cite{kogut1975hamiltonian,robson1982gauge}
\begin{align}\label{eq:KS}
    H_\text{KS}=\frac{g^2}{2a} \sum_{\ell} E_\ell^2 - \frac{1}{2a g^2} \sum_\square \left(\mathcal{U}_\square + \mathcal{U}_\square^\dagger\right) \;,
\end{align}
where $g^2$ is the dimensionless bare coupling constant and $a$ denotes the spatial lattice spacing. Here,  $E_\ell^2$ is the electric energy operator acting on every link $\ell$ of a 2D square lattice, while $\mathcal{U}_\square$ acts on the four links forming an elementary plaquette (see Fig.~\ref{fig1}a). In the Hamiltonian formulation, gauge invariance is expressed by Gauss' law operators $G_+$, associated to every vertex $+$ of the lattice, such that $\left[H_\text{KS},G_+\right] = 0 \; \forall +$, and the gauge-invariant Hilbert space is spanned by all states $|\psi\rangle$ which fulfill Gauss' law $G_+ |\psi \rangle = 0$ (in the absence of static charges).

Since we will define the $q$KS theory in a gauge-invariant basis formed by spin network (SN) states, we first recall this construction for the standard KS model~\cite{robson1982gauge}.
These states are obtained by solving Gauss' law in terms of spin singlets at every four-vertex. To keep track of inequivalent singlets, it is convenient to work on a tri-valent lattice obtained by ``point-splitting'' every four-vertex into two three-vertices as indicated in Fig.~\ref{fig1}b, a construction which is also heavily used in the LSH formulation~\cite{raychowdhury2020loop,raychowdhury2019low}. The fact that this procedure is fundamentally non-unique implies the existence of local basis changes [see Fig.~\ref{fig1}d] which will become essential for SNAQs.
A general SU(2) SN state has the form $|\boldsymbol{j}\rangle = \otimes'_{\ell} |j_\ell\rangle$ with one SU(2) representation label $j_\ell \in \{0, \frac{1}{2}, 1, \dots \}$ assigned to every link of the resulting lattice. The rules of angular momentum addition lead to an additional ``triangle'' constraint $|j_1-j_2| \le j_3 \le j_1 + j_2$, together with $j_1 + j_2 + j_3\in \mathbb{N}$, which has to be satisfied by all triples of spins $(j_1,j_2,j_3)$ that meet at a vertex, which we indicate by the primed product. One can show that the collection of all such SN states forms an orthonormal basis of the gauge-invariant Hilbert space (see \cite{robson1982gauge} and SM). 

We propose to regularize the KS model by deforming the corresponding defining Lie \emph{algebra}. 
In the present example, we proceed by replacing the data arising from the representation theory of SU(2) with analogous expression for the quantum group SU(2)$_k$ (see, e.g., \cite{biedenharn1995quantum} and the SM). 
More precisely, we define generalized SN states with $j_\ell \in \{0, \frac{1}{2}, 1, \dots, \frac{k}{2}\}$, which truncates the local Hilbert space dimension that physically corresponds to a maximum electric flux $j_\text{max} = \frac{k}{2}$. Additionally, the triangle constraint for triples $(j_1, j_2, j_3)$ is replaced by the SU(2)$_k$ fusion rule: $j_1 + j_2 \ge j_3$ and $j_1 + j_2 + j_3 \le k$. 
 To remain close to the original KS model, we define the electric energy operator $E_\ell^2$ analogously and only truncate it to admissible states. That is, $E_\ell^2$ is diagonal and acts only on the links $\ell$ that are also present in the original square lattice (the additional links introduced in the point-splitting do not carry electric energy), where we have $E_\ell^2|j_\ell\rangle = \mathcal{E}(j_\ell) |j_\ell \rangle$ with $\mathcal{E}(j) = j(j+1)$.

To complete our construction, recall that in the SN basis of the KS model, the plaquette operator acts non-trivially on the six inner links of a plaquette, depending on the six outer links (see Fig.~\ref{fig1}c)~\cite{robson1980gauge}. The non-vanishing matrix elements are most conveniently expressed in terms of $F$-matrices (see SM for an explicit formula in terms of Wigner's $6j$-symbols) as
\begin{align}\label{eq:plaquette}
\langle\boldsymbol{j}' |\mathcal{U}_\square|\boldsymbol{j}\rangle &= F^{j_{12}  j_1 j_2 }_{\frac{1}{2} j_2'j_1'}
	 F^{j_{23} j_2 j_3 }_{\frac{1}{2}  j_3'j_2'}
	 F^{j_{34}  j_3 j_4 }_{\frac{1}{2}  j_4'j_3'} \nonumber \\
    &\qquad \times 
	 F^{j_{45}  j_4 j_5 }_{\frac{1}{2}j_5'j_4' }
	 F^{j_{56} j_5 j_6 }_{\frac{1}{2} j_6'j_5' }
	 F^{ j_{61}  j_6 j_1}_{\frac{1}{2} j_1'j_6'},
\end{align}
where a trivial action for other links $\ell$ not touching the plaquette $\square$ is implicit.
For the $q$-deformed theory, we define the action of plaquette operators in the SU(2)$_k$ SN basis by Eq.~\eqref{eq:plaquette} with $F$-matrices replaced by their corresponding counterparts for SU(2)$_k$ (see \cite{biedenharn1995quantum} and the SM).

The resulting theory, which we call the \emph{q-deformed} Kogut-Susskind model ($H_\text{qKS}$), can be interpreted as a particular perturbation of the stringnet models introduced in~\cite{levin2005string}.
A related $q$-deformed truncation of the partition function of 3D SU(2) lattice Yang-Mills theory was studied with tensor network methods in~\cite{cunningham2020tensor}. While the present discussion builds on gauge-covariant bases in the Hamiltonian formulation as introduced in~\cite{robson1982gauge}, note that similar constructions have been used for the LSH formulation~\cite{raychowdhury2020loop}. Gauge-invariant bases have also been constructed for SU(2) quantum link models, enabling efficient Quantum Monte-Carlo simulations through an equivalent dual model~\cite{banerjee2018s} (see also \cite{cherrington2007dual} for a dual formulation of SU(2) lattice Yang-Mills theory).

As we demonstrate in the rest of the paper, the qKS formulation is ideal for simulations with quantum technologies.
In particular, it is constructed such that we recover the KS description of LGTs in the limit $k\rightarrow \infty$ in contrast to, e.g., finite subgroup truncations.  Moreover, the $q$-deformed theory preserves the structure of local unitary transformations of the SN basis in terms of so-called $F$-moves (see Fig.~\ref{fig1}d), which enables a relatively simple decomposition of plaquette operators in contrast to, e.g., quantum link models. This feature also enables the construction of efficient quantum algorithms.

\begin{figure}
        \centering
        \includegraphics[width=\columnwidth]{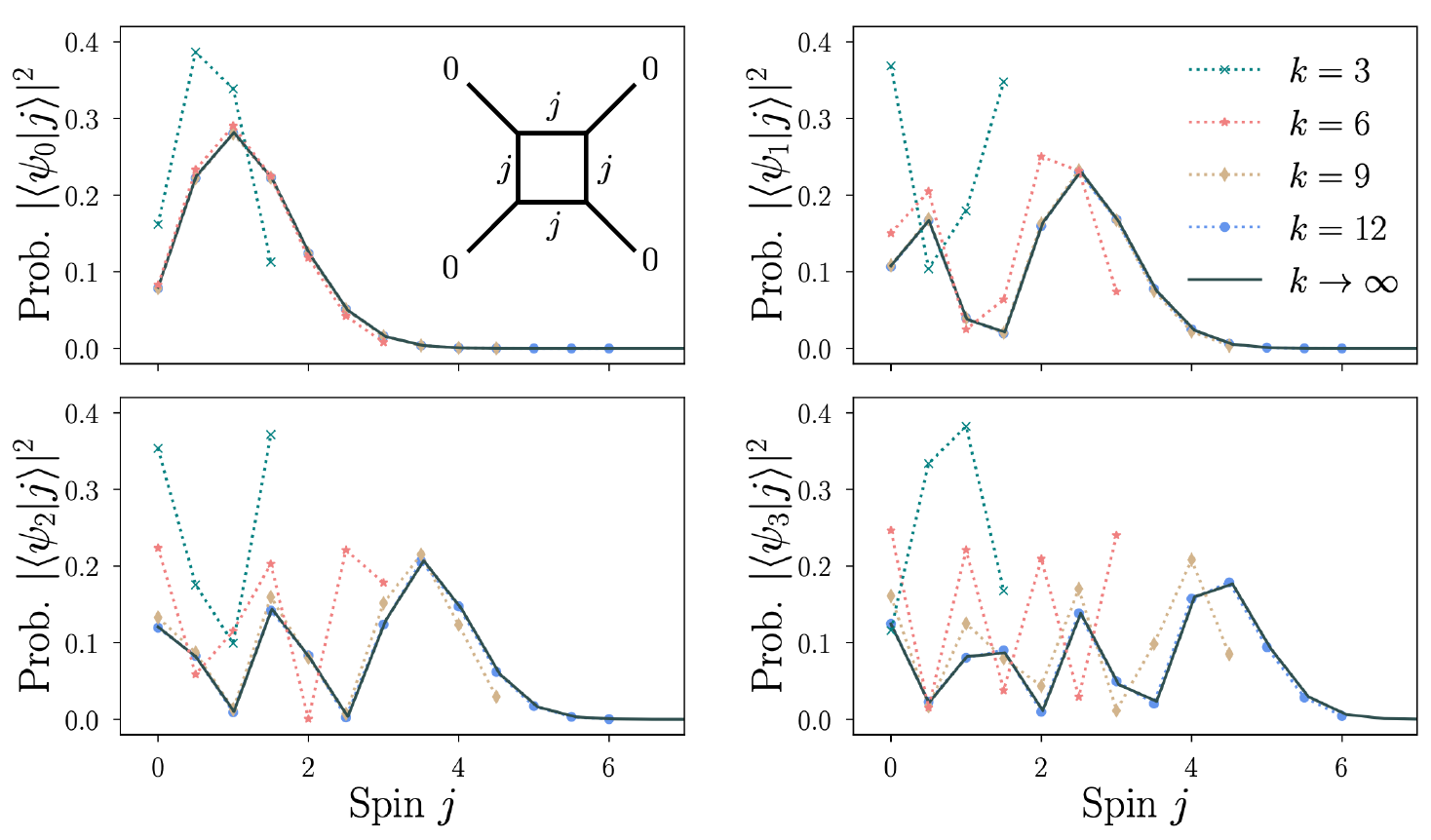}
        \caption{The probability distributions $|\langle \psi|j \rangle|^2$ for the ground state (top left), first (top right), second (bottom left) and third (bottom right) excited states for fixed coupling $g^2 = 0.1$ of a single plaquette converge to the desired $k\rightarrow\infty$ limit quickly once the cutoff $j_\text{max} = \frac{k}{2}$ is large enough to support the bulk of the wavefunction. The inset in the top left panel illustrates the SN basis for a single plaquette with open boundary conditions and zero incoming flux.}
        \label{fig:ED_results}
    \end{figure}

\paragraph{Exact results for a single plaquette.--}
We next illustrate the convergence of our proposed truncation. Consider a single plaquette with
open boundary conditions and fixed zero electric flux at the boundaries as indicated by the SN diagram in the inset of Fig.~\ref{fig:ED_results}. In this case, the gauge-invariant Hilbert space becomes $(k+1)$-dimensional, spanned by SN states $|j\rangle$ with a single label $j$.
	The Hamiltonian (rescaling $H'_{\text{qKS}} = \frac{2a}{g^2} \times H_{\text{qKS}}$) explicitly reads
    \begin{align}
    	H'_{\text{qKS}} &\!=\! \!\sum_{j=0}^{k/2}4\mathcal{E}(j) |j \rangle \langle j|  \!-\! \frac{2}{g^4} \sum_{j=0}^{(k\!-\!1)/2} \!\left(|j\!+\!\tfrac{1}{2} \rangle \langle j| \!+\! \text{h.c.}  \right), 
    \end{align}
    where the effect of working with the generalized SU(2)$_k$ theory is particularly transparent as it just imposes a cutoff $j_\text{max} = \frac{k}{2}$ on the largest flux value allowed on the single plaquette.

    In Fig.~\ref{fig:ED_results}, we plot the probability distributions $|\langle \psi|j \rangle|^2$ corresponding to the ground state $|\psi_0\rangle$, as well as the first three excited states $|\psi_{1/2/3}\rangle$ for a fixed coupling. These results, obtained by exact diagonalization, are compared to the analytical results in terms of Mathieu functions of the limit $k\rightarrow \infty$ (see SM). 
    We observe that the wave-functions converge rapidly for sufficiently large values of $k$, where the threshold is essentially dictated by the total energy and shifts to larger values for higher excited states. Similarly, larger values of $k$ will be need to reach small $g^2$ required for scaling towards the continuum limit.

\paragraph{Classical SNAQ for ground states.--}

The continuum field theory limit is approached by increasing the lattice size and sending $g^2 \rightarrow 0$. While a detailed study of this limit lies beyond the scope of this work, we provide first estimates of how to scale $k$ when decreasing $g^2$ in the following.

To this end, we make a variational ansatz $|\boldsymbol{\psi}\rangle$ for the ground state of an infinite system
	\begin{align}\label{eq:iPEPS_ansatz}
	|\boldsymbol\psi\rangle = \prod_\square\left[\sum_{j=0}^{k/2}\psi_j \mathcal{U}_\square^{(j)} \right]|\mathbf{0}\rangle \;,
	\end{align}
	which is a generalization of the one used in Refs.~\cite{dusuel2015mean,vanderstraeten2017bridging}.
	Here, $|\mathbf{0}\rangle$ is the SN vacuum state, $\mathcal{U}^{(j)}_\square$ is the plaquette operator that creates a $j$ flux loop on the plaquette $\square$, i.e. replacing $1/2$ by $j$ in Eq.~\eqref{eq:plaquette}. The $\psi_j$ are variational parameters, which
    are normalized as $\sum_{j=0}^{k/2} |\psi_j|^2 = 1$.

	There are several reasons for using this ansatz: First, it can exactly represent ground states in the limiting cases $g^2 = 0$ and $g^2\rightarrow\infty$. Second, as shown in the SM, we can evaluate the expectation value of $H_{q\text{KS}}$ analytically and find
	\begin{align}\label{eq:mean_energy}
	    \langle \boldsymbol\psi| H'_\text{qKS} | \boldsymbol\psi \rangle &\propto \sum_{j_1 j_2 j_3} |\psi_{j_1}|^2 |\psi_{j_2}|^2 \frac{j_3 (j_3 +1)d_{j_3}}{d_{j_1}d_{j_2}} \delta_{j_1 j_2 j_3} 
	    \nonumber\\
	    &\qquad -\frac{1}{g^4} \sum_{j_1 j_2} \psi_{j_1}^* \psi_{j_2} \delta_{j_1 j_2 \frac{1}{2}} \;.
	\end{align}
	Here, $\delta_{j_1 j_2 j_3}$ abbreviates the fusion constraint that $(j_1 j_2 j_3)$ form an admissible vertex and $d_j$ is the quantum dimension of $j$ (see SM for details ). We want to emphasize that even though the ansatz has a ``mean-field-like'' character, it in general represents a highly entangled state. Technically, it can be interpreted as an 
    iPEPS (see also \cite{gu2009tensor,buerschaper2009explicit,tagliacozzo2014tensor,vanderstraeten2017bridging,robaina2021simulating}). We expect that generalizations of this tensor network ansatz will be very useful for future investigations with classical high-performance computing, as well as with quantum hardware, or hybrid variational approaches.

    \begin{figure}
        \centering
        \includegraphics[width=\columnwidth]{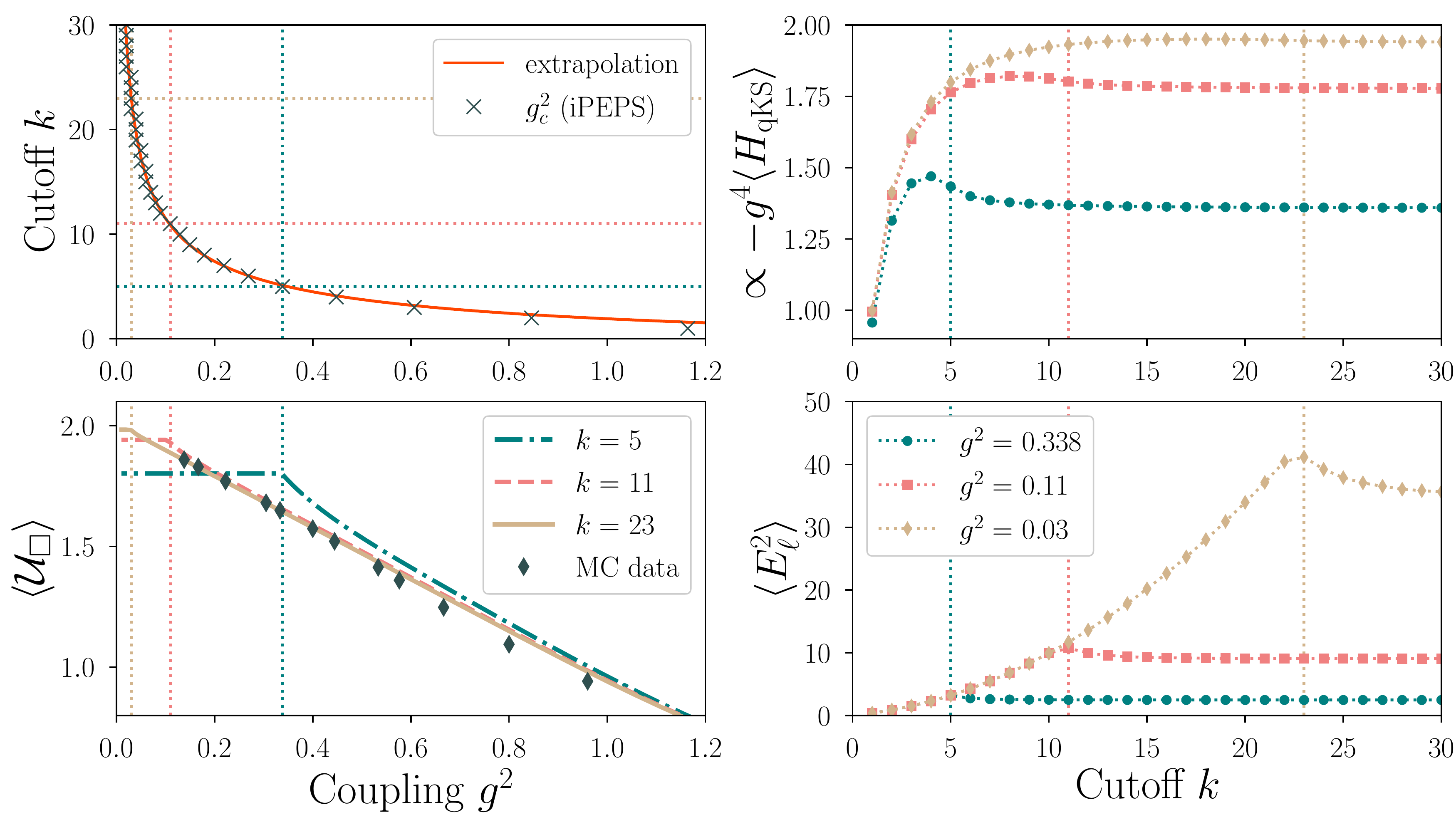}
        \caption{The top left panel shows the critical coupling $g_c^2$ as a function of $k$, extracted from the non-analytic behaviour of the mean plaquette $\langle \mathcal{U}_\square \rangle = \langle \boldsymbol\psi_\text{opt}| \mathcal{U}_\square| \boldsymbol\psi_\text{opt} \rangle$ (lower left panel) in the optimized iPEPS $|\boldsymbol\psi_\text{opt}\rangle$. Both the total energy (top right), as well as the local electric energy (bottom right) converge rapidly with increasing $k$, once the threshold $k_c$ is surpassed.
        The dotted vertical and horizontal lines indicate the relation between the values of $k_c$ and $g_c^2$. We compare our iPEPS results for $\langle \mathcal{U}_\square\rangle$ with MC data taken from Table VIII of Ref.~\cite{teper1998n}.
        \label{fig:iPEPS}}
    \end{figure}

    Here, we find 
    an approximation of the ground state as a function of $g^2$ for several $k$ by numerically minimizing the average energy [Eq.~\eqref{eq:mean_energy}]. Our results are summarized in Fig.~\ref{fig:iPEPS}. For large $g^2$, the system is in a confined phase as expected for a strong electric field energy, which is also the phase expected for the continuum theory~\cite{svetitsky1982critical}. For generic finite values of $k$, however, we observe indications of a phase transition for small $g^2$. For $k=1$, this phase is expected to be topologically ordered, i.e.~deconfined, with $\mathbb{Z}_2$ (Toric code) topological order~\cite{levin2005string}.
    Note that the undesired phases (from a high-energy physics point of view) shrink towards $g^2 \rightarrow 0$ as $k$ is increased.

    As illustrated in Fig.~\ref{fig:iPEPS}, we find fast convergence of local observables with increasing $k$ once the system is in the anticipated ``correct'' phase. 
    This further motivates us to consider the location $g_c^2=g_c^2(k)$ of the transition as an estimate for the value $k_c = k_c(g^2)$ when the model significantly deviates from the desired continuum behavior. For given a coupling $g^2$, we expect to converge to the continuum limit rapidly for $k\gtrsim k_c(g^2)$. Our findings are consistent with a simple power-law behavior of the form $g_c^2 = \left(\frac{g_0}{k+k_0}\right)^2$ with $g_0 \approx 4.4$
    and $k_0\approx 2.5$,
    which agrees with the expectation that $g_c^2 \rightarrow 0$ as $k\rightarrow \infty$.
    This suggests that a moderately small coupling like $g^2 = 0.1$
    requires $k = g_0/ g - k_0 \sim \mathcal{O}(10)$, which lies within reach of trapped-ion qudit computers~\cite{ringbauer2022universal} by encoding a single link into a single qudit.
    
    In practice, it is sufficient to decrease the coupling $g^2$ until the \emph{scaling regime} is reached, where the continuum physics can be reliably extracted.
    For the $2+1$D SU(2) KS model, we compare our simulations to Euclidean Monte-Carlo (MC) results for the plaquette expectation value~\cite{teper1998n}.
    We can obtain quantitative agreement with the MC data in the regime $k\gtrsim
    15$
    and $0.1 \lesssim g^2 \lesssim 0.5$,
    indicating that our tensor network ansatz -- despite its simplicity -- captures the essential degrees of freedom correctly.

\paragraph{Quantum SNAQ for real-time evolution.--}
To illustrate the usefulness of our proposed formulation for quantum simulation, we now present a quantum SNAQ that provides an exact Trotter decomposition of the time-evolution operator of the $q$-deformed theory.
The algorithm is formulated on a \emph{SN register}, where we associate one degree of freedom $|j_\ell\rangle$ to every link $\ell$ of the hexagonal graph obtained from point-splitting the original lattice. We will refer to $|j_\ell\rangle$ as a local qudit, but a further decomposition into qubits is of course possible. Note that this computational basis is overcomplete because it contains states violating the fusion constraints.
We keep this redundancy here because it simplifies gate parallelization within the SNAQ, thus making the approach scalable to large system sizes. Furthermore, since the constraints imposed by the fusion rules are diagonal in the computational basis, configurations that do not correspond to valid SN states can be dealt with relatively easily.

	The core elements of this SNAQ are local basis changes ($F$-moves), which in particular allow diagonalizing the plaquette operator.  On the SN register, an $F$-move corresponds to a multiply-controlled unitary operator that changes the state of one target qudit, depending on the state of four control qudits (see Fig.~\ref{fig1}d) as
	\begin{align}\label{eq:F_operator}
	    F|j_1 j_2 j_3 j_4 j\rangle = |j_1 j_2 j_3 j_4 j'\rangle \;,
	\end{align}
    where the operator $F$ is defined by the matrix elements $\left(F^{j_1 j_2}_{j_3 j_4}\right)_{j,j'} = F^{j_1 j_2 j'}_{j_3 j_4 j}$. 
    This five-qudit operator $F$ induces other controlled unitaries with less controls. Explicitly, we will need a four-qudit operator $F'$ which is defined through the matrix elements $\left({F'}^{j_1}_{j_3 j_4}\right)_{j,j'} = F^{j_1 j_1 j'}_{j_3 j_4 j}$, identifying $j_1=j_2$.
    Finally, we introduce a controlled two-qudit operator $G$, which diagonalizes the matrix $\left(F''_{J}\right)_{j'j} = F^{J j j}_{\frac{1}{2} j'j'}$ whose eigenvalues we denote by $\omega^{(J)}_j$.

    	\begin{figure*}
		\centering
    \includegraphics[width=2\columnwidth]{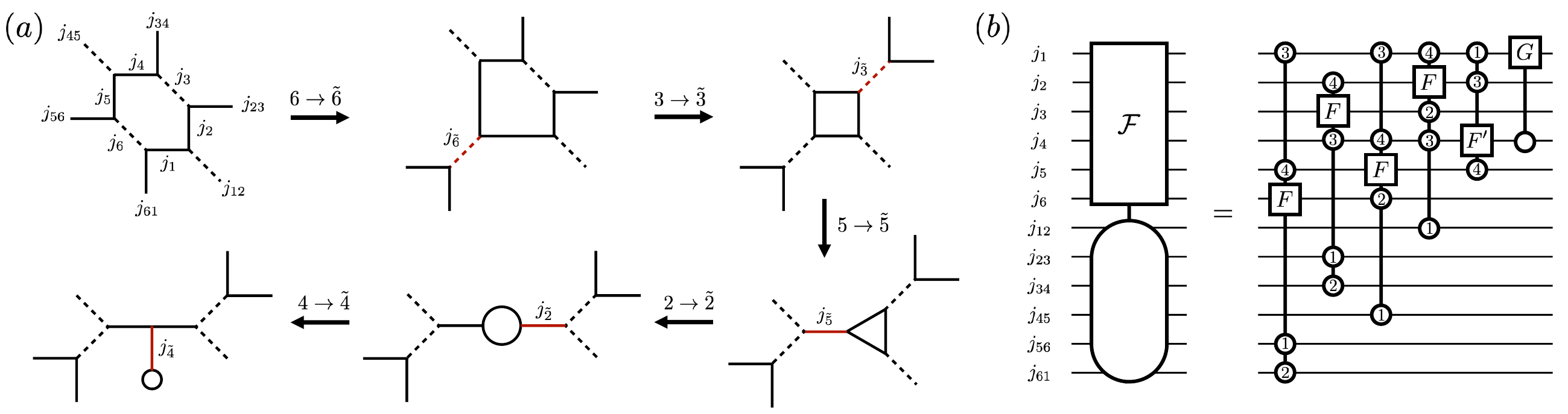}
		\caption{\label{fig:plaq_diag_moves} (a) Sequence of five $F$-moves that partially diagonalizes the plaquette term for a hexagon-shaped SN. Dashed lines indicate auxiliary links that arise from the virtual point-splitting of four-vertices on a 2D square lattice. The links which are affected by a single $F$-move are highlighted in red in the resulting SN diagram. For example, the first $F$-move denoted as $6 \rightarrow \tilde{6}$ involves the links $j_{61}, j_1, j_5, j_{56}$ and changes $j_6$ to $j_{\tilde{6}}$.
    (b) Quantum circuit decomposition derived from the local unitary transformation illustrated in (a).}
	\end{figure*}

    Now, we arrive at a key observation:
    There exists a sequence of $F$-moves, shown in Fig.~\ref{fig:plaq_diag_moves}a, that partially diagonalizes the plaquette operator on an elementary hexagon. Intuitively, the properties of $F$-matrices allow to shrink the loop of the plaquette down the elementary matrix element $F^{\tilde{j_4} j_1 j_1}_{\frac{1}{2} j_1'j_1'}$ (see SM for details). To the best of our knowledge, this property was first observed in~\cite{robson1982gauge} for the original KS theory and later translated to quantum circuits for stringnet models~\cite{bonesteel2012quantum}. Our proposed $q$-deformed regularization is tailored to preserve this property.
    As a direct consequence, we obtain the controlled unitary quantum circuit $\mathcal{F}$ shown in figure~\ref{fig:plaq_diag_moves}b. The operator $\mathcal{F}$ acts on the inner qudits of a hexagon $j_1, \dots, j_6$ and takes the outer qudits $j_{12}, \dots, j_{61}$ as controls.
    This decomposition enables an analytic way to deal with the plaquette operator -- made possible by the unitarity of $F$-moves, a property that is lost in other formulations -- which we expect will be beneficial in many quantum algorithms for LGTs.
    In SM we further provide explicit decompositions of the involved unitaries into controlled two-qudit gates, demonstrating a simple and transparent implementation on a \emph{qudit} quantum computer~\cite{gonzalez2022hardware,ringbauer2022universal}.
    
    An immediate application is a 
    SNAQ based on an analytical Trotter decomposition of the evolution operator $U_{\text{qKS}}(\tau) = e^{-i\tau H_{\text{qKS}}}$. Explicitly, we write a Trotter step of a single plaquette term as
    \begin{align}
        e^{i\tau \frac{2}{ag^2}  \mathcal{U}_\square} \!=\! \mathcal{F} \Omega(\tau) \mathcal{F}^\dagger  \,, \;\Omega(\tau)|j_1 j_4 \rangle \!=\! e^{i\tau \frac{2}{ag^2} \omega^{(j_4)}_{j_1}}|j_1 j_4 \rangle
    \end{align}
    where $\Omega(\tau)$ denotes the appropriate two-qudit phase gate. For a 2D square lattice this Trotter step can be applied in parallel on half of all plaquettes, 
    yielding an exact realization of the magnetic part of the time evolution operator, $U_B (\tau) = e^{+i\tau \frac{2}{ag^2} \sum_\square\mathcal{U}_\square} = \prod_\square e^{+i\tau \frac{2}{ag^2} \mathcal{U}_\square}$. The electric part $U_E (\tau) = e^{-i\tau \frac{g^2}{2a} \sum_\ell E_\ell^2} = \prod_\ell e^{-i\tau \frac{g^2}{2a} E_\ell^2}$ can be trivially parallelized in terms of single-qudit phase gates $e^{-i\tau \frac{g^2}{2a} E_\ell^2}|j_\ell \rangle = e^{-i\tau \frac{g^2}{2a} j_\ell(j_\ell +1)}|j_\ell \rangle$ on all physical links. $U_{\text{qKS}}$ can then be approximated in terms of $U_B$ and $U_E$ as usual.
    
    As an example, let us briefly analyze the required quantum resources of a standard second-order algorithm $U_{\text{qKS}}(\tau)  =  U_{E}(\tau/2) \times U_{B}(\tau) \times U_{E}(\tau/2) + \mathcal{O}(\tau^2)$. For this purpose, we 
    assume that every link is encoded into a single qudit of size $k+1$.
    Including possible parallelizations, a single Trotter step has a circuit depth determined by $2$ electric phase gates ($E_\ell^2$), $2$ magnetic phase gates ($\Omega$), $4$ applications of $G$ and $F'$, as well as $12$ full $F$ gates. We quantify the circuit complexity $C$ by the number of controlled two-qudit unitaries using the decompositions shown in SM, which yields a polynomial scaling of $C \le 4 + 28(k+1)^3 + 108 (k+1)^4 \sim \mathcal{O}(k^4)$. Using the properties of the $F$-matrices, we expect that the exact gate count can be drastically improved, and leave further optimizations for future work.

\paragraph{Outlook.--}
    Our work sets the stage for several follow-up investigations. First, an extension to general SU($N$), in particular $N=3$, gauge theories is desirable. In this case, a technical obstacle will be the book-keeping of multiplicities in the generalized Clebsch-Gordon series, which could be overcome using a graphical calculus as developed in~\cite{hamer1986cluster,liegener2016towards} adapted to the $q$-deformed case. Second, it appears straightforward to incorporate matter, either fermionic or Higgs fields, into our approach, which will add some matter-specific gates to the SNAQ~\cite{https://doi.org/10.48550/arxiv.2303.06985, https://doi.org/10.48550/arxiv.2303.08683}. Third, given the close similarities to the spin-foam approach to quantum gravity~\cite{rovelli2015covariant,dittrich2016decorated}, it will be interesting to explore related classical and quantum simulations of gravity~\cite{asaduzzaman2020tensor,cohen2021efficient}.
     From a condensed matter perspective, the $q$-deformed KS LGTs proposed here deserve further study in their own right as interesting topologically-ordered phases~\cite{levin2005string} and novel types of critical phenomena~\cite{somoza2021self} can be expected, and we refer to ~\cite{koenig2010quantum,bonesteel2012quantum,liu2022methods} for related methods to simulate anyons on a quantum computer.
	
    	Classically, we expect that the use of gauge-invariant tensor networks~\cite{tagliacozzo2014tensor,dittrich2016decorated,vanderstraeten2017bridging,magnifico2021lattice,meurice2022tensor,emonts2023finding} will play a crucial role in simulations of gauge theories. For the gauge theories studied in this work, extensions of the ansatz in Eq.~\eqref{eq:iPEPS_ansatz} to inhomogeneous or time-dependent scenarios could be particularly useful to study the dynamics of (de)confined flux strings, as well as string breaking.
    On the quantum side, existing and near-future quantum hardware, in particular based on qudits~\cite{wang2020qudits,ringbauer2022universal,gonzalez2022hardware,https://doi.org/10.48550/arxiv.2303.06985,https://doi.org/10.48550/arxiv.2303.08683}, provide
    the means for implementing the algorithm presented here, as well as other quantum or hybrid variational SNAQs. 

\paragraph{Acknowledgements.}
T.V.Z. thanks Luca Tagliacozzo for illuminating discussions concerning tensor network representations of gauge-invariant states. The authors thank Z. Davoudi, I. Raychowdhury and J. R. Stryker for discussions about the LSH formulation, and
J. Dziarmaga for careful reading of the manuscript.
This work was supported by the Simons Collaboration on Ultra-Quantum Matter, which is a grant from the Simons Foundation (651440, P.Z.)

\bibliography{references}

\clearpage

\appendix
\onecolumngrid
\section*{Supplemental Material to ``Quantum and classical spin network algorithms \\for $q$-deformed Kogut-Susskind gauge theories''}

In this supplemental material, we provide detailed calculations and list known facts that have been omitted in the main text for brevity. In particular, we review essential facts about SU(2)$_k$ and the gauge-invariant SN formulation of SU(2) LGT, including the analytic solution of the single plaquette case. Morevoer, we show how to locally diagonalize the plaquette operator using $F$-moves, calculate observables used for the iPEPS simulation presented in the main text, and discuss explicit gate decompositions for SNAQs.
\\

\twocolumngrid

\section{Spin network basis for SU(2)}
The SN basis for 2+1D SU(2) LGT was first discussed in~\cite{robson1982gauge}, which we briefly review in this section.
The full Hilbert space associated to a 2D square lattice is spanned by states $\otimes_\ell |j_\ell m_\ell n_\ell\rangle$. For every link $\ell$, we have left- and right-electric fields $\mathbf{L}_\ell$ and $\mathbf{R}_\ell$, which are spin-operators with the same length, $E_\ell^2 = \mathbf{R}_\ell^2 = \mathbf{L}^2_\ell$, corresponding to the basis states $|j_\ell m_\ell n_\ell\rangle = |j_\ell m_\ell\rangle  |j_\ell n_\ell\rangle$.
Let us focus on a single vertex, formed by four links $1,2,3,4$, where the 
Gauss' law operator is given by \mbox{$\mathbf{G} = \mathbf{L}_1 + \mathbf{L}_2 + \mathbf{R}_3 + \mathbf{R}_4$}. Finding all state $|\psi\rangle$ that are gauge-invariant states, $\mathbf{G}|\psi\rangle = 0$ therefore boils down to constructing singlets out of the basis states $|j_1 m_1\rangle |j_2 m_2\rangle |j_3 n_3\rangle |j_4 n_4\rangle$. Following the usual rules of angular momentum addition, we first fuse $1,2$ and $3,4$ and then add the result to obtain the states
\begin{widetext}
\begin{align}
|j_1 j_2 j_3 j_4 j_{12} \rangle = \sum_{m_1, m_2, n_3, n_4} \frac{(-1)^{j_{12}-m_{12}}}{\sqrt{2j_{12}+1}} C^{j_{12},m_{12}}_{j_1, m_1, j_2, m_2} C^{j_{12},-m_{12}}_{j_3, n_3, j_4, n_4}|j_1 m_1\rangle |j_2 m_2\rangle |j_3 n_3\rangle |j_4 n_4\rangle  \;,
\end{align} 
\end{widetext}
where $C^{J,M}_{j_1,m_1,j_2,m_2}$ are SU(2) Clebsch-Gordan coefficients. This construction clearly shows the origin of the ``point-splitting'' and the fusion constraints discussed in the main text. Repeating this procedure for all vertices yields the SN basis. Here, different angular momentum addition schemes yield inequivalent bases. In particular, locally fusing $2,3$ and $4,1$ first leads to another way of labelling all local singlets, which we denote by $|j_1 j_2 j_3 j_4 j_{12}\rangle$. Their overlap is defines the $F$-matrix of SU(2),
\begin{widetext}
\begin{align}\label{eq:F_SU(2)}
F^{j_1 j_2 j_{12}}_{j_3 j_4 j_{23}} = \langle j_1 j_2 j_3 j_4 j_{23}|j_1 j_2 j_3 j_4 j_{12} \rangle = (-1)^{j_1 + j_2 + j_3 + j_4} \sqrt{(2j_{12}+1)(2j_{23}+1)} \begin{Bmatrix}
    j_1 & j_2 & j_{12}\\
    j_3 & j_4 & j_{23}
\end{Bmatrix}  \;,
\end{align}
\end{widetext}
where the curly bracket denotes Wigner's $6j$ symbol. To obtain the matrix elements of the KS Hamiltonian, note that $E_\ell^2$ is already diagonal with eigenvalues $j_\ell(j_\ell+1)$. We omit the calculation of matrix elements of the plaquette operator, which is  most conveniently performed using the graphical calculus developed  in~\cite{yutsis1962mathematical}. The form presented in the main text follows from the results of ~\cite{robson1982gauge} upon using \eqref{eq:F_SU(2)}. 

\section{Facts about SU(2)$_k$\label{app:SU(2)k}}
	 For completeness, we list facts about SU(2)$_k$ in this section, largely following Sec. III. A and appendix A of \cite{dittrich2017quantum}, as well as \cite{kirillow1989representations}.

    In the following $q = e^{2\pi i/(k+2)}$ denotes a fixed root of unity with $k$ a positive integer.
    Several expressions in this section look identical for SU(2)$_k$ and the familiar case of SU(2)~\cite{messiah1962quantum}, which essentially follows from the Schwinger boson construction~\cite{biedenharn1989quantum}, replacing ordinary numbers $n$ and factorials $n!$ by their $q$-deformed analogs $[n](!)$, 
     Here, the $q$-factorial $[n]! = [n] [n-1] \cdots [1]$ is defined in terms of the $q$-number
	\begin{align}\label{eq:q_number}
		[n] = \frac{q^{n/2}-q^{-n/2}}{q^{1/2}- q^{-1/2}} = \frac{\sin\left(\frac{\pi }{k+2}n\right)}{\sin\left(\frac{\pi }{k+2}\right)}
	\end{align}
 	and $[0]! = [0] = 1$. 
  
	 The SU(2)$_k$ fusion rule
	 \begin{align}
	     j_1 \times j_2 = \sum_{j_3} \delta_{j_1 j_2 j_3} j_3
	 \end{align}
	 determines how two representations labelled by $j_1$ and $j_2$ can be recoupled to $j_3$. Here, we
	 use the notation
	\begin{align}\label{eq:fusion_delta}
	    \delta_{j_1 j_2 j_3} = \begin{cases}
	    1 \;,  & (j_1, j_2, j_3) \quad \text{admissible} \\
	    0 \;, & (j_1, j_2, j_3) \quad \text{not admissible}
	    \end{cases} 
	\end{align}
	for admissible triples $(j_1, j_2, j_3)$, which satisfy the fusion constraints
	\begin{subequations}
	\begin{align}
	    j_1 + j_2 \ge j_3 \;,\\
	    j_2 + j_3 \ge j_1 \;,\\
	    j_3 + j_1 \ge j_2 \;,\\
	    j_1 + j_2 + j_3 \le k \;,\\
	    j_1 + j_2 + j_3 \in \mathbb{N} \;.
	\end{align}
	\end{subequations}
	We have the symmetry $\delta_{j_1 j_2 j_3} = \delta_{j_2 j_3 j_1} = \delta_{j_3 j_1 j_2}$, and
	$j=0$ is the unit element of fusion, i.e. $\delta_{j_1 0 j_3} = \delta_{j_1 j_3}$. Moreover, fusion is associative,
	\begin{align}
	    \sum_{j} \delta_{j_1 j_2 j} \delta_{j j_3 j_4} = \sum_j \delta_{j_1 j_4 j} \delta_{j j_2 j_3} \;.
	\end{align}
	
	 The $F$-matrices of SU(2)$_k$ can be defined analogous to the SU(2) case as
	 \begin{align}\label{eq:F-matrices}
	     F^{j_1 j_2 j_5}_{j_3 j_4 j_6} = (-1)^{j_1 + j_2 + j_3 + j_4}\sqrt{d_{j_5} d_{j_6}} \begin{Bmatrix}
	         j_1 & j_2 & j_5\\
	         j_3 & j_4 & j_6
	     \end{Bmatrix} \;,
	 \end{align}
	  where $d_j = [2j+1]$ is the quantum dimension of $j$ and the curly bracket now denotes the \emph{$q$-deformed} $6j$ symbol, given by the Racah formula
	  \begin{widetext}
	  \begin{align}\label{eq:Racah}
	      \begin{Bmatrix}
	         j_1 & j_2 & j_5\\
	         j_3 & j_4 & j_6
	     \end{Bmatrix}  &= \sqrt{\Delta_{j_1 j_2 j_5} \Delta_{j_1 j_4 j_6} \Delta_{j_3 j_2 j_6} \Delta_{j_3 j_4 j_5}} \,\sum_j (-1)^j [j+1]!  \\
	     &\times\frac{\left( [j_1 + j_2 +j_3 + j_4 -j]! \, [j_1 + j_3 +j_5 + j_6 - j]! \, [j_2 + j_4 + j_5 + j_6 -j]! \right)^{-1}}{[j- j_1 - j_2 - j_5]! \, [j- j_1 - j_4 - j_6]! \, [j - j_3 - j_2 - j_6]! \, [j- j_3 - j_4 - j_6]!} \;. \nonumber
	\end{align}
	Here, the sum runs over all integers $j$ satisfying $\text{max} \le j \le \text{min}$ with
	\begin{align}
	    \text{max} &= \max \left(j_1 + j_2 + j_5, \, j_1 + j_4 + j_6, \, j_3 + j_2 + j_6, \, j_3 + j_4 + j_5\right) \;,\\
	    \text{min} &= \min \left(j_1 + j_2 + j_3 + j_4 , \, j_1 + j_3 + j_5 + j_6 , \, j_2 + j_4 + j_5 + j_6\right) \;.
	\end{align}
	and we abbreviated
	\begin{align}
	    \Delta_{j_1 j_2 j_3} = \delta_{j_1 j_2 j_3} \frac{[j_1 + j_2 - j_3]! \, [j_1 - j_2 + j_3]! \, [-j_1 + j_2 + j_3]!}{[j_1 + j_2 + j_3 + 1]!} \;.
	\end{align}
	\end{widetext}
	
	To remember these expressions, it is useful to associate the $6j$ symbols to a tetrahedron formed by four triangles $(j_1, j_2, j_5)$, $(j_1, j_4, j_6)$, $(j_3, j_2, j_6)$, $(j_3, j_4, j_5)$, where every $j$ labels one side of a triangle. By definition, the $6j$ symbols vanish unless the four triples are admissible. The $6j$ symbol enjoys the associated tetrahedral symmetry under any permutation of indices in the columns and pair-wise exchange of row-indices. For the $F$-matrices, this symmetry translates to
	\begin{align}
	    F^{j_1 j_2 j_5}_{j_3 j_4 j_6} = F^{j_2 j_1 j_5}_{j_4 j_3 j_6} = F^{j_4 j_3 j_5}_{j_2 j_1 j_6} = F^{j_5 j_2 j_1}_{j_6 j_4 j_3} \frac{v_{j_5} v_{j_6}}{v_{j_2} v_{j_4}} \;,
	\end{align}
	where $v_j^2 = (-1)^{2j} d_j$. All $F$-matrices are real, $\left(F^{j_1 j_2 j_5}_{j_3 j_4 j_6}\right)^* = F^{j_1 j_2 j_5}_{j_3 j_4 j_6}$ and are normalized such that 
	\begin{align}
	    F^{j_1 j_1 0}_{j_2 j_2 j_3} = \frac{v_{j_3}}{v_{j_1} v_{j_2}} \delta_{j_1 j_2 j_3} \;.
	\end{align}
	Moreover, the $6j$ symbols obey a version of the Biedenharn-Elliot identity, often called pentagon identity for the $F$-matrices,
	\begin{align}\label{eq:pentagon}
	\sum_J F^{j_1 j_2 j_5}_{j_3 j_4 J} F^{j_6 j_7 j_4}_{J j_1 j_8} F^{j_8 j_7 J}_{j_3 j_2 j_9} = F^{j_1 j_2 j_5}_{j_9 j_6 j_8} F^{j_6 j_7 j_4}_{j_3 j_5 j_9} \;.
	\end{align}
	Finally, we have the orthogonality relation
	\begin{align}\label{eq:orthogonality}
	    \sum_{J} F^{j_1 j_2 j'}_{j_3 j_4 J} F^{j_1 j_2 j}_{j_3 j_4 J} = \delta_{j' j} \;.
	\end{align}

 \section{Diagonalization of an elementary plaquette operator}
 In this section, we prove that local $F$-moves partially diagonalize the plaquette operator by shrinking the size of the involved loop. Consider the first move illustrated in the main text, which affects the neighboring corners of the link $j_6$. The involved matrix element is transformed as
\begin{align}
\sum_{j_6'j_6} &F^{j_{56} j_{61} \tilde{j}_6'}_{j_1'j_5'j_6'} F^{j_{61} j_6 j_1}_{\frac{1}{2} j_1'j_6'} F^{j_{56} j_5 j_6}_{\frac{1}{2} j_6' j_5'} F^{j_{56} j_{61} \tilde{j}_6}_{j_1 j_5 j_6}\\
&=\sum_{j_6} \frac{v_{j_1} v_{j_6} v_{j_5'}}{v_{j_5} v_\frac{1}{2} v_{j_6'}} F^{j_6 j_{56} j_5}_{\tilde{j}_6' j_1 j_{61}} F^{j_1 j_1'\frac{1}{2}}_{j_{5}' j_{5} \tilde{j}_6'} F^{j_{56} j_{61} \tilde{j_6}}_{j_1 j_5 j_6}\\
&= \delta_{\tilde{j}_6 \tilde{j}_6'} F^{\tilde{j}_6 j_5 j_1}_{\frac{1}{2} j_1' j_5'},
\end{align}
where we have used the symmetries of $F$-matrices to re-order the indices, and applied the pentagon identity and the orthogonality relation in the first and second lines, respectively. This calculation shows that the structure of the plaquette operator remains unchanged under $F$-moves. It simply shrinks (or grows) and consists of one $F$-matrix per corner in the loop.
Performing also the other steps, we obtain the transformed matrix element in the new SN basis,
\begin{align}
\langle \boldsymbol{\tilde{j}}'| \mathcal{U}_\square | \boldsymbol{\tilde{j}} \rangle = \delta_{\tilde{j}_6'\tilde{j}_6} \delta_{\tilde{j}_5'\tilde{j}_5}
\delta_{\tilde{j}_4'\tilde{j}_4} 
\delta_{\tilde{j}_3'\tilde{j}_3} 
\delta_{\tilde{j}_6'\tilde{j}_2} 
F^{\tilde{j}_4 j_1 j_1}_{\frac{1}{2} j_1' j_1'} \;,
\end{align}
where we again omitted the implicit Kronecker deltas of all other indices outside of the original plaquette.

\section{Single Plaquette}
The results of the previous sections also imply that the plaquette operator of the single plaquette case discussed in the main text is determined by
\begin{align}
\langle j' | \mathcal{U}_\square | j \rangle = \left(F^{0 j j }_{\frac{1}{2} j'j'} \right)^4 = \delta_{j j'\frac{1}{2}} \;.
\end{align}
The full Hamiltonian $H'_\text{qKS}$ thus becomes a simple tri-diagonal matrix with entries $4j(j+1)$ on the diagonal, and $-\frac{2}{g^4}$ on the first off-diagonal.

We now provide an analytical solution for the limit $k \rightarrow \infty$. Following~\cite{robson1980gauge}, we change to a basis given by $|x\rangle = \sum_j \sin \left[(2j+1)x\right] |j\rangle$ with $x\in [0,\pi)$. The Hamiltonian then takes the form
\begin{align}
H'_\text{KS} = -\left(\partial_x^2 +1\right) - \frac{4}{g^4} \cos(x) \;,
\end{align}
such that the eigenvalue problem is solved in terms of Mathieu functions. In our case, the relevant eigenfunctions $\psi(x) = \langle \psi|x \rangle$ must be $2\pi$-periodic and anti-symmetric, i.e. $\psi(x+2\pi) = \psi(x)$ and $\psi(-x) = -\psi(x)$. The corresponding solutions are the sine-elliptic functions $\psi_n(x) = se_{2n+2}\left(q,\frac{x}{2}\right)$ with $n=1,2,3, \dots$ and $q=\frac{1}{2g^4}$, which have the well-known Fourier expansion
\begin{align}
se_{2n+2}\left(q,\frac{x}{2}\right) = \sum_{j} B^{(2n+2)}_{4j+2}(q) \sin \left[\left(2j+1\right)x \right] \;,
\end{align}
where the sum runs over half-integers. In the SN basis $|j\rangle$, the eigenfunctions $|\psi_n\rangle$ are thus determined by the Mathieu coefficients $B_{4j+2}^{(2n+2)} = \langle \psi_n | j \rangle$, which are shown for the numerical comparison discussed in the main text.

\section{Calculation of the mean energy in the iPEPS ansatz}
In this section, we calculate observables for the ansatz
\begin{align}
	|\boldsymbol\psi\rangle = \prod_\square\left[\sum_{j=0}^{k/2}\psi_j \mathcal{U}_\square^{(j)} \right]|\mathbf{0}\rangle \;.
\end{align}
We will rely on results of \cite{levin2005string}, where it was shown that the operators $\mathcal{U}_\square^{(j)}$ on different plaquettes commute and on same plaquettes obey
\begin{align}
\mathcal{U}_\square^{(j_1)} \mathcal{U}_\square^{(j_2)} = \sum_{j_3} \delta_{j_1 j_2 j_3} \mathcal{U}_\square^{(j_3)}
\end{align}
when applied to a spin network state.
First note that the variational ansatz is properly normalized,
\begin{align}
\langle \boldsymbol{\psi} | \boldsymbol{\psi} \rangle &= \prod_{\square \square'} \sum_{j_1,j_2} \psi_{j_1} \psi^*_{j_2} \langle \mathbf{0}|\mathcal{U}_\square^{(j_1)} \mathcal{U}_{\square'}^{(j_2)}|\mathbf{0} \rangle \\
&= \prod_\square \sum_{j_1} \left|\psi_{j_1}\right|^2  = 1 \;,
\end{align}
where we used that $\langle \mathbf{0}| \mathcal{U}_\square^{(j)} | \mathbf{0}\rangle = \delta_{j0}$.

Similarly, the expecation value of a single plaquette operator $\mathcal{U}_\square = \mathcal{U}^{\left(\frac{1}{2}\right)}_\square$ becomes
\begin{align}
\langle \boldsymbol{\psi} |\mathcal{U}_\square| \boldsymbol{\psi} \rangle &= \sum_{j_1 j_2} \psi_{j_1} \psi_{j_2}^* \langle \mathbf{0}|\mathcal{U}_\square^{(j_1)} \mathcal{U}^{\left(\frac{1}{2}\right)}_\square\mathcal{U}_{\square}^{(j_2)}|\mathbf{0} \rangle \\
&= \sum_{j_1 j_2} \psi_{j_1} \psi_{j_2}^* \delta_{j_1 j_2 \frac{1}{2}} \;.
\end{align}

The calculation of the electric energy $E_\ell^2$ is more lengthy as it involves two plaquettes, which denote by $\square$ and $\square'$, adjacent to the link $\ell$. The expectation value takes the form
\begin{align}
\langle \boldsymbol{\psi} |E_\ell^2| \boldsymbol{\psi} \rangle &= \sum_{j_1 j_2 j_3 j_4} \psi_{j_1}^* \psi_{j_2}^* \psi_{j_3} \psi_{j_4} \mathcal{M}^{(j_1 j_2 j_3 j_4)}_\ell  \;,
\end{align}
where we abbreviated
\begin{widetext}
\begin{align}
\mathcal{M}^{(j_1 j_2 j_3 j_4)}_\ell &= \langle \mathbf{0}| \mathcal{U}^{(j_1)}_\square \mathcal{U}^{(j_2)}_{\square'} E_\ell^2 \mathcal{U}^{(j_3)}_\square \mathcal{U}^{(j_4)}_{\square'} | \mathbf{0}\rangle\\
&= \sum_{abcdehijklg\tilde{f}f} g(g+1) \, F^{000}_{j_1 \tilde{f} a}F^{000}_{j_1 a b} F^{000}_{j_1 b c} F^{000}_{j_1 c d} F^{000}_{j_1 d e} F^{000}_{j_1 e \tilde{f}} F^{a \tilde{f}0}_{j_2 l g}  F^{e0\tilde{f}}_{j_2 g h}  F^{000}_{j_2 k i} F^{000}_{j_2 i j} F^{000}_{j_2 j k} F^{000}_{j_2 k l}\nonumber\\
&\qquad\qquad\qquad\qquad\qquad\times F^{000}_{j_1 \tilde{f} a}F^{000}_{j_1 a b} F^{000}_{j_1 b c} F^{000}_{j_1 c d} F^{000}_{j_1 d e} F^{000}_{j_1 e \tilde{f}} F^{a \tilde{f}0}_{j_2 l g}  F^{e0\tilde{f}}_{j_2 g h}  F^{000}_{j_2 k i} F^{000}_{j_2 i j} F^{000}_{j_2 j k} F^{000}_{j_2 k l}\\
&=\delta_{j_1 j_4} \delta_{j_2 j_3} \sum_g g(g+1) F^{j_1 j_1 0}_{j_2 j_2 g} F^{j_1 0 j_1}_{j_2 g j_2} F^{j_4 j_4 0}_{j_3 j_3 g} F^{j_4 0 j_4}_{j_3 g j_3} = \delta_{j_1 j_4} \delta_{j_2 j_3} \delta_{j_1 j_2 g} \sum_g g(g+1) \frac{d_g}{d_{j_1} d_{j_2}} \;.
\end{align}
In this calculation, we have used the normalization of the $F$-matrices and the special case $F^{000}_{abc} = \delta_{ab} \delta_{bc}$.
Collecting everything, we find the average energy density
\begin{align}
\frac{1}{N_\square}\left\langle \boldsymbol{\psi} \bigg|\sum_\ell E_\ell^2  - \frac{2}{g^4} \sum_\square \mathcal{U}_\square\bigg|  \boldsymbol{\psi} \right\rangle &= 2 \sum_{j_1 j_2 j_3} \left|\psi_{j_1}\right|^2 \left|\psi_{j_2}\right|^2 \frac{j_3 (j_3+1) d_{j_3}}{d_{j_1} d_{j_2}} \delta_{j_1 j_2 j_3} -  \frac{2}{g^4}  \sum_{j_1j_2} \psi^*_{j_1} \psi_{j_2} \delta_{j_1 j_2 \frac{1}{2}} \;,
\end{align}
\end{widetext}
where $N_\square\rightarrow\infty$ is the number of all plaquettes. To get the right prefactors, we used the fact that only $2N_\square$ of all $3N_\square$ links carry electric energy.

\section{Explicit gate decompositions}
In this section, we consider further decompositions of the multi-qudit gates introduced in the main text into elementary single qudit gates $\mathcal{U}$ and controlled two-qudit gates $C\mathcal{U}$. The gate set $\{\mathcal{U}, C\mathcal{U}\}$ can be directly realized, e.g., with the architecture put forward in~\cite{gonzalez2022hardware} or further decomposed into native gates demonstrated in recent experiments with trapped ions~\cite{ringbauer2022universal}.

Consider a general $n$-controlled qudit operator $C_{j_1, \dots, j_n}^{(n)}\mathcal{U}$ that acts with single-qudit unitary $\mathcal{U}$ onto the $(n+1)$ qudit if and only if the first $n$ qudits are in particular states $j_1, \dots, j_n$, i.e.
\begin{align}
C_{j_1, \dots, j_n}^{(n)}\mathcal{U} &= \left(\bigotimes_{\ell=1}^{n} \mathbf{1} - \bigotimes_{\ell=1}^{n} |j_\ell\rangle  \langle j_\ell| \right) \otimes \mathbf{1} \nonumber \\ &\qquad+ \left(\bigotimes_{\ell=1}^{n} |j_\ell\rangle  \langle j_\ell| \right) \otimes \mathcal{U} \,.
\end{align}
A five-qudit $F$-move gate can be decomposed into a product of at most $(k+1)^4$ such gates as
\begin{align}
F = \prod_{j_1,j_2,j_3,j_4}C^{(4)}_{j_1,j_2,j_3,j_4}\mathcal{U}_{F}
\end{align}
for an appropriate choice of $\mathcal{U}_{F}$.
Similarly the four-qudit gate $F'$ can be decomposed into at most $(k+1)^3$ $3$-controlled gates. Including arbitrary single-qudit rotations the problem thus reduces to constructing the gate $C^{(n)}\mathcal{U} \equiv C^{(n)}_{0, \dots, 0}\mathcal{U}$.

\begin{figure}[b]
    \centering
    \includegraphics[width=0.8\columnwidth]{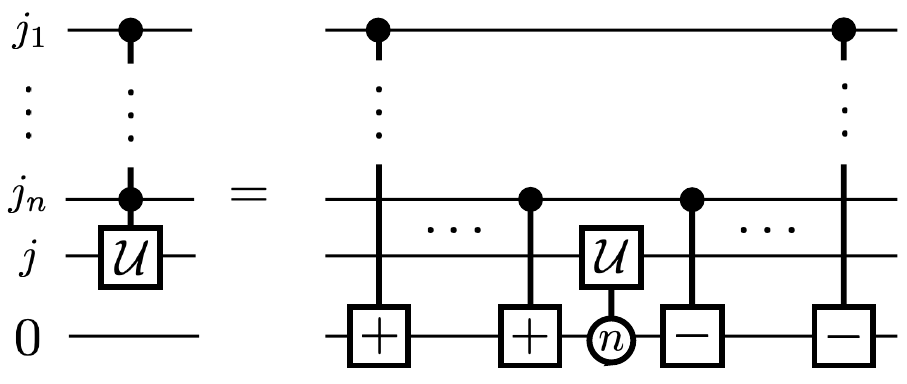}
    \caption{Decomposition of an $n$-controlled qudit gate $C^{(n)}\mathcal{U}$ into $2n+1$ $1$-controlled qudit gates using one ancilla qudit of size $\ge n$.}
    \label{fig:multiply-controlled}
\end{figure}

Following~\cite{muthukrishnan2000multivalued}, we decompose $C^{(n)}\mathcal{U}$ using ancilla qudits. In fact, a single ancilla of size $n+1$ is sufficient, which fits nicely into our proposal since $n\le 4$ and the natural qudit size is $k+1$. A simple possiblity is to realize $C^{(n)}\mathcal{U}$ is illustrated in Fig.~\ref{fig:multiply-controlled}. Here, the control condition of the first $n$ qudits is first written into the ancilla using $n$ controlled two-qudit addition gates. The desired operation is then realized by a single controlled two-qudit gate, and aftewards the ancilla is disentangled again. The complexity of this protocol is given by $2n+1$ basic entangling gates.

\end{document}